\documentclass[pra,twocolumn,showpacs,showkeys,amsmath,amssymb]{revtex4}
\pdfoutput=1
\usepackage{graphicx}
\usepackage{dcolumn}
\usepackage{bm}

\begin{document}
\title{All-Optical Switching with Transverse Optical Patterns}
\author{Andrew M. C. Dawes}
\author{Lucas Illing}
\author{Joel A. Greenberg}
\author{Daniel J. Gauthier}
\email{gauthier@phy.duke.edu}
\affiliation{
Department of Physics, the Center for Nonlinear and Complex Systems, and the 
Fitzpatrick Institute for Photonics and Communication Systems, Duke University, 
Durham, North Carolina 27708
}
\date{\today}

\begin{abstract} 
We demonstrate an all-optical switch that operates at ultra-low-light levels 
and exhibits several features necessary for use in optical switching networks. 
An input switching beam, wavelength $\lambda$, with an energy density of 
$10^{-2}$ photons per optical cross section [$\sigma=\lambda^2/(2\pi)$] changes 
the orientation of a two-spot pattern generated via parametric instability in 
warm rubidium vapor. The instability is induced with less than 1~mW of total 
pump power and generates several $\mu$Ws of output light. The switch is 
cascadable: the device output is capable of driving multiple inputs, and 
exhibits transistor-like signal-level restoration with both saturated and 
intermediate response regimes. Additionally, the system requires an input power 
proportional to the inverse of the response time, which suggests thermal 
dissipation does not necessarily limit the practicality of optical logic 
devices.
\end{abstract}



\maketitle

\section{Introduction} 
\label{sec:intro}

Optical switches are a crucial component of communication networks where light 
is redirected from channel-to-channel \cite{Gibbs_1985aa} and in general 
computational machines where they can act as logic elements 
\cite{Keyes_2006aa}. For all-optical switches, where light controls the flow of 
light, there has been a continual push to increase the sensitivity of the 
switch so that it can be actuated with lower powers and hence decreasing the 
system complexity. With the advent of quantum information systems, it is 
important to increase the sensitivity to the point where a single switching 
photon is effective \cite{Bouwmeester_2000aa}.

In a recent brief communication \cite{Dawes_2005aa}, we described a novel 
approach for achieving ultra-low-light-level all optical switching. This 
approach involves combining near-resonance, sub-Doppler nonlinear optics with 
optical pattern formation. In particular, we observe that patterns generated by 
a counterpropagating-beam instability can be switched with a pulse of light 
that has less than 3,000 photons. The primary purpose of this paper is to give 
additional details about our experiment and to extend the work to: demonstrate 
and characterize the transistor-like action of the switch, peform a detailed 
study of the time response of the switch, and to study the pattern-forming 
instability as various experimental parameters are varied.

The paper is structured as follows. In the next section, we review the basic 
requirements for a switch with classical and quantum information applications, 
describe previous approaches for all-optical low-light-level switching, and 
review the mechanisms that give rise to optical pattern formation. Section 
\ref{sec:setup} describes our experimental system, while 
Sec.~\ref{sec:instability} details our observations of the light generated by 
the pattern-forming instability. In Sec.~\ref{sec:switching}, we present the 
results of several measurements used to characterize the response of our switch 
to various input power levels. Finally, Sec.~\ref{sec:discussion} contains a 
discussion and analysis of these results.


\section{Background} 
\label{sec:background}

Switches can be used in two classes of applications: information networks and 
computing systems. In each of these applications, information can be stored in 
either classical or quantum degrees of freedom. Hence, the requirements for a 
device vary depending on the intended application.

Classical, all-optical networks require switches to reliably redirect or gate a 
signal depending on the presence of a control field at the device input. 
Ideally, the switch shows large contrast between on and off output levels and 
can be actuated by low input powers. If the network carries quantum 
information, the switch must be triggered by an input field containing only a 
single quanta (photon). Additionally, the quantum state of the transmitted 
signal field must be preserved.

If a switch is to be used as a logic element in a classical computing system, 
it must have the following characteristics: input-output isolation, 
cascadability, and signal level restoration \cite{Keyes_2006aa}. Input-output 
isolation prohibits the device output from having back-action on the device 
input. Cascadability requires that a device output have sufficient power to 
drive the input of at least two identical devices. Signal level restoration 
occurs in any device that outputs a standard signal level in response to a wide 
range of input levels. That is, variations in the input level do not cause 
variations in the output level. Switching devices that satisfy these 
requirements are considered scalable devices, \emph{i.e.}, the properties of 
the individual device are suitable for scaling from one device to a network of 
many devices.

While scalability describes important properties of a switching device, 
sensitivity provides one way to quantify its performance. A highly sensitive 
all-optical switch can be actuated by a very weak optical field. Typical 
metrics for quantifying sensitivity are: the input switching energy (in 
Joules), the input switching energy density (in photons per optical cross 
section $\sigma=\lambda^2/2\pi$, where $\lambda$ is the wavelength of the input 
beam) \cite{Harris_1998aa, Keyes_1970aa}, and the total number of photons in 
the input switching pulse.

One may not expect a single device to satisfy all of the requirements for these 
different applications. For example, a switch operating as a logic element 
should output a standard level that is insensitive to input fluctuations. This 
may be at odds with quantum switch operation where the device must preserve the 
quantum state of the signal field. An interesting question arises from these 
requirements: What happens when a classical switch is made sensitive enough to 
respond to a single photon? Reaching the level of single photon sensitivity has 
been the goal of a large body of recent work that is reviewed below.

\subsection{Previous Research on Low-Light-Level Switching}

Two primary approaches to low-light-level switching have emerged, both of which 
seek to increase the strength of the nonlinear coupling between light and 
matter. The first method uses fields and atoms confined within, and strongly 
coupled to, a high-finesse optical cavity. The second method uses traveling 
waves that induce quantum interference within an optical medium and greatly 
enhance the effects of light on matter.


Cavity quantum-electrodynamic (QED) systems offer very high sensitivity by 
decreasing the number of photons required to saturate the response of an atom 
that is strongly coupled to a mode of the cavity. Working in the 
strong-coupling regime, Hood \emph{et al.}\ measured the transmission of a 
10~pW probe beam through a cavity with linewidth $\kappa=40$~MHz (cavity 
lifetime 25~ns) while cold caesium atoms were dropped through the cavity mode. 
When an atom is present in the cavity mode, the effect of a single cavity 
photon (on average), is an order of  magnitude increase in cavity transmission. 
For 10 cavity photons, the nonlinear optical response is saturated and almost 
complete transmission is observed \cite{Hood_1998aa}. Operating with a cavity 
mode waist of 15~$\mu$m, a single input photon ($\lambda=852$~nm) represents an 
input energy density of $\sim$$10^{-4}$~photons/$\sigma$ and a total input 
energy of $\sim$$10^{-19}$~J.

In a similar system, Birnbaum \emph{et al.}\ \cite{Birnbaum_2005aa} observe an 
effect known as photon-blockade, where the absorption of a single photon 
prevents subsequent absorption of a second photon. For the probe frequency used 
in these experiments, the single-photon process is resonant with the lowest 
excited dressed-state of the atom-cavity system while the two-photon absorption 
process is suppressed. With an average of 0.21~photons in the cavity (mode 
waist $w=23.4 \mu$m), the sensitivity of this system is 
$\sim$$10^{-5}$~photons/$\sigma$, comparable to the lowest reported to date 
\cite{Zhang_2007aa}.

Although a single two-level atom in free space also exhibits an effect similar 
to photon blockade: once excited it cannot immediately absorb a second photon, 
interactions between single photons and single atoms in free space are 
exceedingly difficult to control. Hence one major achievement of cavity QED is 
to control the single-photon, single-atom regime. The primary drawback to 
integrating cavity-QED-based devices into switching networks is that the cavity 
system is designed to operate in a single field mode, limiting the input and 
output channels to one per polarization. One must then discriminate between 
signal photons and control photons in some way other than by input mode (such 
as by polarization). 

Of the scalability requirements for an all-optical switch, cavity QED systems 
do not strictly satisfy cascadability, where the output of one device must be 
capable of driving two ore more subsequent inputs. While coupling light to an 
atom contained in the cavity is efficient, all input and output signals are 
coupled strongly. Thus, the input and the output must have similar powers, 
otherwise the stronger beam completely overrides the effect of the weaker beam. 
While cavity QED systems are well suited for controlling one single-photon 
signal with another single photon, they are not designed to allow single-photon 
inputs to control strong, many-photon signals.

A different technique for all-optical switching in cavities relies on creating 
and controlling cavity solitons. The most recent experiments use vertical 
cavity surface emitting lasers (VCSELs) as the nonlinear cavity 
\cite{Hachair_2005aa}. A VCSEL can be prepared for cavity solitons by injecting 
a wide holding beam along the cavity axis the cavity. A narrow write beam 
traveling through the laser cavity induces a cavity soliton. Typically, the 
solitons persist until the holding beam is turned off, hence this system 
naturally serves as an optical memory.

Lower sensitivity is the primary limitation of the cavity soliton systems. 
Typical powers for the hold and write beams are 8~mW and 150~$\mu$W 
respectively. The lowest reported write beam power is 10~$\mu$W for a holding 
beam of 27~mW, suggesting a compromise can be made to lower the required write 
power by increasing the hold beam power. Injecting 10~$\mu$W during the 500~ps 
turn-on time corresponds to an input pulse containing $\sim$24,000~photons, and 
a switching energy density of $\sim$140 photons/$\sigma$ (write beam diameter 
10~$\mu$m, $\lambda=960-980$~nm). Extinguishing, or switching, solitons 
involves either cycling the holding beam or injecting a second ``write'' pulse 
out of phase to erase the soliton, thus adding a slight complication to the 
device. 

In terms of the scalability criteria discussed above, cavity soliton devices 
satisfy the requirements of signal level restoration because the soliton 
intensity stabilizes to a consistent level. Unlike cavity QED devices, 
large-area cavity soliton systems are inherently multi-mode and can be made 
very parallel with each soliton location serving as an isolated  channel. To be 
cascadable, however, each cavity soliton would have to emit enough power to 
seed solitons in two or more subsequent devices. Additional techniques may then 
be required to properly image the output soliton into a second cavity. This 
problem has not been addressed in the literature to the best of our knowledge.


Traveling wave approaches can also operate with multi-mode optical fields and 
achieve few-photon sensitivity. Recent progress has been made through the 
techniques of electromagnetically induced transparency (EIT) 
\cite{Harris_1997aa,Schmidt_1996aa,Zibrov_1999aa,Braje_2003aa, 
Chen_2005aa,Zhang_2007aa}. As an example, Harris and Yamamoto 
\cite{Harris_1998aa} proposed a switching scheme using the strong 
nonlinearities that exist in specific states of four-level atoms where, in the 
ideal limit, a single photon at one frequency causes the absorption of light at 
another frequency. To achieve the lowest switching energies, the narrowest 
possible atomic resonances are required, which is the main challenge in 
implementing this proposal.

Using the narrow resonances offered by trapped cold atoms, and the 
Harris-Yamamoto scheme, Braje \emph{et al.}\ \cite{Braje_2003aa} first observed 
all-optical switching in an EIT medium with an input energy density of 
$\sim$23~photons/$\sigma$. Subsequent work by Chen \emph{et al.}\ 
\cite{Chen_2005aa} confirmed that such EIT switching operates at the 
1~photon/$\sigma$ level. Using a modified version of the Harris-Yamamoto scheme 
with an additional EIT coupling field that causes additional quantum 
interference, Zhang \emph{et al.}\ \cite{Zhang_2007aa} recently observed 
switching with $\sim $20 photons ($10^{-12}$~W for $\tau=$0.7~$\mu$s with a 
0.5~mm beam diameter) corresponding to $10^{-5}$~photons/$\sigma$. This is the 
highest all-optical switching sensitivity reported to date.

Although EIT switches are very sensitive, the input and output fields are 
necessarily of the same strength so the requirements for cascadability are not 
met. Additionally, the output level for EIT based switches is a monatonically 
decreasing function of input level \cite{Braje_2003aa}, thus the output level 
is sensitive to variations in the input level and the switches do not perform 
signal level restoration.

Other low-light-level all-optical switching experiments have also been 
demonstrated recently in traveling-wave systems. By modifying the correlation 
between down-converted photons, Resch \emph{et al.}\ \cite{Resch_2002aa} 
created a conditional-phase switch that operates at the single photon level. 
Using six-wave mixing in cold atoms, Kang \emph{et al.}\ \cite{Kang_2004aa} 
demonstrated optical control of a field with 0.2 photons/$\sigma$ with a 
2~photon/$\sigma$ input switching field ($\sim$$10^8$ input photons, over 
$\sim$0.54~$\mu$s in a $\sim$0.5~mm diameter beam). Both of these results 
exhibit high sensitivity but, like the EIT schemes, they are limited to control 
fields that are stronger than the output field.

Another approach combines the field enhancement offered by optical cavities 
with the strong coupling of coherently prepared atoms. Bistability in the 
output of a cavity filled with a large-Kerr, EIT medium \cite{Wang_2002aa} 
exhibits switching, but requires higher input power: 
$\sim$$10^8$~photons/$\sigma$ (0.4~mW, 80~$\mu$m radius, $\sim$2~$\mu$s 
response time). Photonic crystal nanocavities have also shown bistability 
switching, where Tanabe \emph{et al.}\ \cite{Tanabe_2005aa} demonstrated 
switching with 74~fJ pulses and a switching speed of $<100$~ps ($\sim$500,000 
photons). Simulations of photonic crystal microcavities filled with an 
ultra-slow-light EIT medium \cite{Soljacic_2005aa} suggest switching could be 
achieved with less than 3000 photons. Taking a different approach, Islam 
\emph{et al.}\ \cite{Islam_1988aa} exploit a modulational instability in an 
optical fiber interferometer to gate the transmission of 184~mW by injecting 
only 4.4~$\mu$W (2,000 photons during the 50~psec switching time). With an 
effective area of 2.6$\times10^{-7}$~cm$^2$, this sensitivity corresponds to 
24~photons/$\sigma$.

Of the two most sensitive systems just discussed, EIT-filled photonic crystal 
microcavities suffer from the same drawbacks as cold-atom EIT systems: the 
input and output fields are required to have the same power, making them not 
cascadable. The other highly sensitive system, a modulational-instability fiber 
interferometer, is both cascadable and exhibits signal level restoration. In 
several ways the latter system is similar to ours: it exploits the sensitivity 
of instabilities and uses a sensitive detector (in their case the 
interferometer, in our case pattern orientation) to distinguish states of the 
switch.

Finally, in a very recent proposal, Chang \emph{et al.}\ \cite{Chang_2007ab} 
suggest that a system consisting of a nano-wire coupled to a dielectric 
waveguide could be used to create a single photon optical transistor. The 
absorption of a single photon by a two-level emitter placed close to the 
nanowire is sufficient to change the nanowire from complete plasmon reflection 
to complete plasmon transmission. This system is similar in effect to cavity 
QED systems, with the added advantage that the input and output are separate 
modes and thus separate channels. If implemented as proposed, a surface-plasmon 
transistor could operate with single-photon input levels, and gate signals 
containing many photons, thus indicating such a device would be cascadable. 

Many all-optical switches have been successfully demonstrated over a period 
spanning several decades. However, in almost every case, one or more important 
features is missing from the switching device. With the requirements of 
scalability and sensitivity in mind, we report a new approach to all-optical 
switching.

\subsection{Pattern formation}

Our approach to all-optical switching is to exploit collective instabilities 
that occur when laser beams interact with a nonlinear medium 
\cite{Lugiato_1994aa}. One such collective instability occurs when laser beams 
counterpropagate through an atomic vapor. In this configuration, given 
sufficiently strong nonlinear interaction strength, it is known that 
mirror-less parametric self-oscillation gives rise to stationary, periodic, or 
chaotic behavior of the intensity \cite{Silberberg_1984aa,Khitrova_1988aa} 
and/or polarization \cite{Gaeta_1987aa,Gauthier_1988aa,Gauthier_1990aa}.

Another well-known feature of counterpropagating beam instabilities is the 
formation of transverse optical patterns, \emph{i.e.}, the formation of spatial 
structure of the electromagnetic field in the plane perpendicular to the 
propagation direction \cite{Petrossian_1992aa,Lugiato_1994aa}. This is also 
true for our experiment where a wide variety of patterns can be generated, 
including rings and multi-spot off-axis patterns in agreement with previous 
experiments \cite{Petrossian_1992aa,Grynberg_1988aa,Gauthier_1990aa}.

Building an all-optical switch from transverse optical patterns combines 
several well-known features of nonlinear optics in a novel way. Near-resonance 
enhancement of the atom-photon coupling makes our system sensitive to weak 
optical fields. Using optical fields with a counterpropagating beam geometry 
allows for interactions with atoms in specific velocity groups leading to  
sub-doppler nonlinear optics without requiring cold atoms. Finally, using the 
different orientations of a transverse pattern as distinct states of a switch  
allows us to maximize the sensitivity of the pattern forming instability. 
Instabilities, by nature, are sensitive to perturbations, so, by combining 
instabilities with resonantly-enhanced, sub-Doppler nonlinearities, we created 
a switch with very high sensitivity.


\section{Experimental Setup} 
\label{sec:setup}

\begin{figure}
\includegraphics{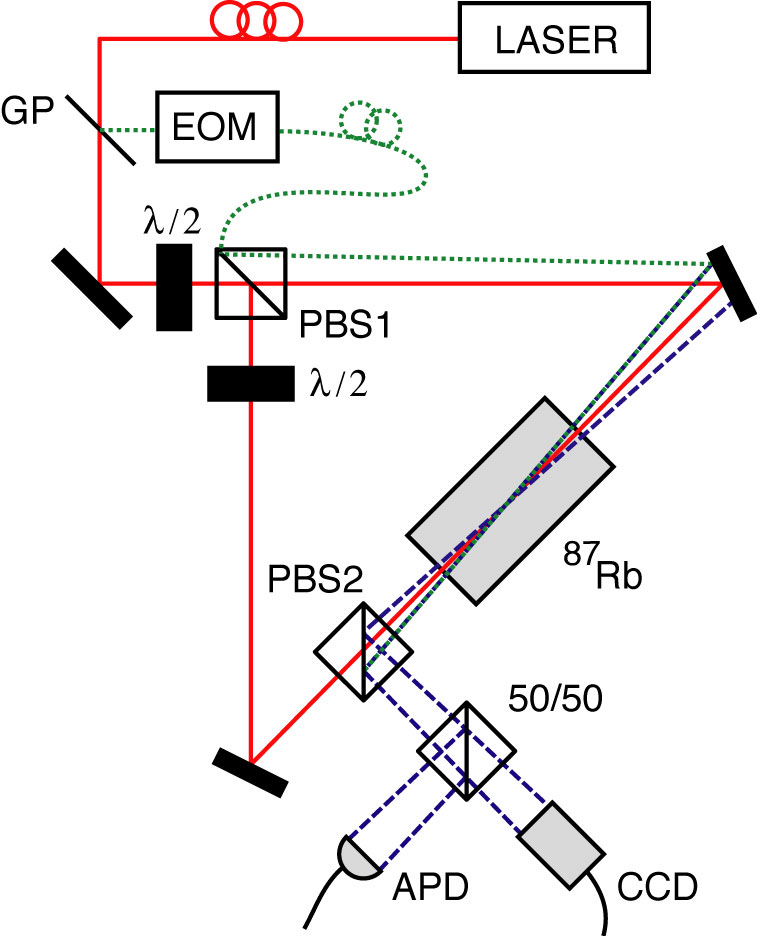}
\caption{\label{fig:setup} Experimental setup for transverse optical pattern 
generation. A cw Ti:Sapphire laser serves as the light source. Optical elements 
are as follows: uncoated glass plate (GP), electro-optic modulator (EOM), 
half-wave plates ($\lambda$/2), polarizing beamsplitters (PBS1, PBS2), 50/50 
beamsplitter (50/50), CCD camera (CCD), and avalanche photodiode (APD). Beams 
are indicated by line type: pump beams (solid), switch beam (dotted), 
instability-generated light (dashed). The foward (backward) pump beam 
propagates in the cw (ccw) direction through the triangular ring cavity.}
\end{figure}

Previous observations of pattern-forming instabilities have required several to 
hundreds of milliwatts of optical power due to the typically weak nonlinear 
interactions and the correspondingly high threshold for self-oscillation. To 
lower the instability threshold, we tune the pump laser close to an atomic 
resonance, and use a pump beam polarization configuration that gives rise to 
polarization instabilities which are known to have lower self-oscillation 
thresholds \cite{Gaeta_1993aa}. The low instability threshold reduces the power 
required to generate optical patterns and gives rise to patterns that exhibit 
sensitivity to perturbation by a weak probe beam, as discussed in 
Sec.~\ref{sec:switching} and Ref. \cite{Dawes_2005aa}.

A diagram of our experimental setup is shown in Fig.~\ref{fig:setup}. Two beams 
of light from a common laser source counterpropagate through warm rubidium 
vapor contained in a glass cell. The light source is a frequency-stabilized 
continuous-wave Ti:Sapphire laser, the output of which is spatially filtered 
using a single-mode optical fiber with an angled entrance face and a 
flat-polished exit face. The beam is then roughly collimated to a spot size 
(1/e field radius) of $w$ = 340 $\mu$m with the beam waist located at the 
center of the vapor cell. The power ratio between the pump beams is controlled 
by a half-wave plate at the input of the first polarizing beam splitter (PBS1). 
We denote the beam passing through PBS1 as the forward beam and the reflected 
beam as the backward beam. A second half-wave plate in the backward beam path 
rotates the polarization such that the pump beams are linearly polarized with 
parallel polarizations.

The atomic medium is isotopically-enriched rubidium vapor ($>$ 90\% $^{87}$Rb), 
which is contained in a 5-cm-long glass cell heated to 67 $^{\circ}$C 
(corresponding to an atomic number density of $\sim 2 \times 10^{11}$ 
atoms/cm$^3$). The cell is tilted with respect to the incident laser beams to 
prevent possible oscillation between the uncoated windows. The cell has no 
paraffin coating on the interior walls that would prevent depolarization of the 
ground-state coherence, nor does it contain a buffer gas that would slow 
diffusion of atoms out of the pump laser beams. The Doppler-broadened linewidth 
of the transition at this temperature is $\sim$550 MHz. To prevent the 
occurrence of magnetically-induced instabilities and reduce Faraday rotation, 
we use a Helmholtz coil to cancel the ambient magnetic field component along 
the direction of the counterpropagating laser beams.

A polarizing beam splitter (PBS2) placed in the beam path separates light 
polarized orthogonally to the pump beam. This light, henceforth referred to as 
\emph{output} light, is subsequently split with a 50/50 beamsplitter and then 
observed simultaneously using any two of the following: a CCD-camera (Marshall 
V-1050A), an avalanche photodiode (Hamamatsu C5460), or a photomultiplier tube 
(Hamamatsu H6780-20) as shown in Fig.~\ref{fig:setup}.

For measurements of the switch response, as discussed in 
Sec.~\ref{sec:switching}, we inject a weak beam at a small angle to the pump 
beams. This switch beam is indicated in Fig.~\ref{fig:setup} by a dotted line. 
An uncoated glass plate reflects $\sim4$\% of the pump beam power into a fiber 
coupler and through a high-speed fiber-based Mach-Zehnder amplitude modulator 
(EOSpace Lithium Niobate Modulator, AZ-2K1-20-PFU-SFU-770, 20 GHz bandwidth). 
Neutral density filters placed before the EOM are used to reduce the switch 
beam power to the range $0.1-1$~nW. The switch beam then enters PBS1 near the 
pump beams, propagates in the forward direction at a small angle ($<5$~mrad) to 
the forward pump beam, and crosses both pump beams in the center of the vapor 
cell.


\section{Characteristics of the Instability-Generated Light} 
\label{sec:instability}

\begin{figure}
\includegraphics{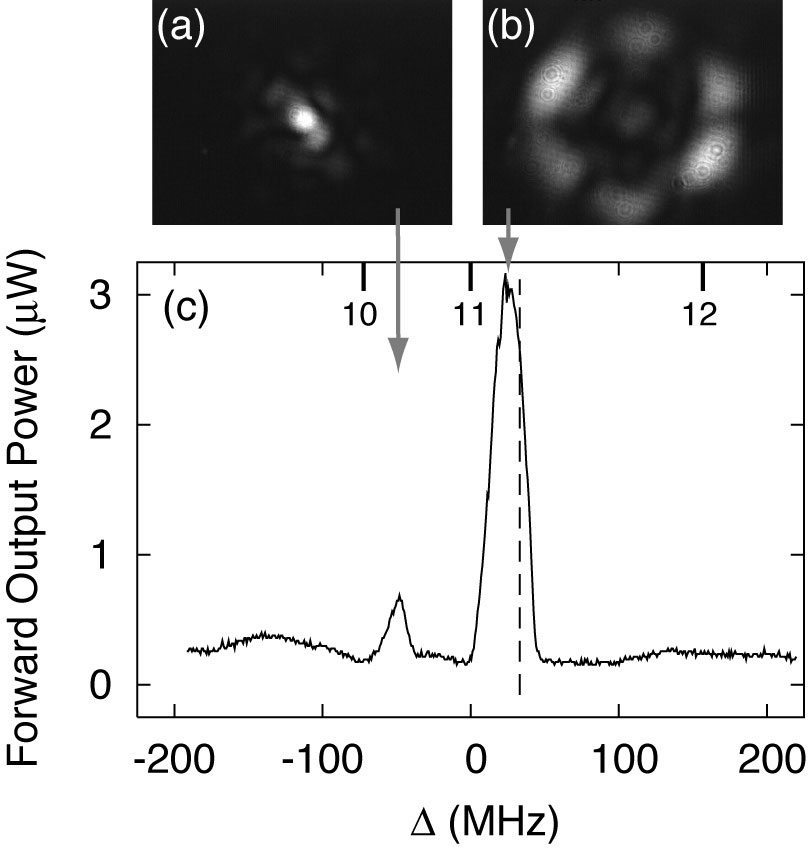}
\caption{\label{fig:output} Instability-generated patterns and optical power as 
a function of pump laser frequency detuning ($\Delta = \nu - \nu_{F=1,F'=1}$). 
(a,b) Patterns for red and blue pump laser detunings ($\Delta=-50$~MHz and 
$\Delta=+25$~MHz respectively, indicated by arrows). (c) Orthogonally polarized 
output power, emitted in the forward direction, as a function of frequency. 
These data correspond to a single scan through the $^5\text{S}_{1/2}(F=1) 
\rightarrow {^5\text{P}_{3/2}}(F')$ transition in $^{87}$Rb from low to high 
frequency. The bold tick marks indicate the hyperfine transitions labeled by 
FF', where F (F') is the ground (excited) state quantum number. Our switching 
experiments are conducted with the pump laser detuned $\Delta=+$30 MHz (dashed 
line). Pump beam power levels: 630 $\mu$W forward and 225 $\mu$W backward.}
\end{figure}

In our setup, the counterpropagating pump beams give rise to an instability 
mechanism that generates new light when the pump power is above a certain 
threshold. For our experimental setup, we observe instability generated light 
(output light) in the state of polarization orthogonal to that of the pump 
beams and with the same frequency as the pump beams.

We find that the power of the output light is maximized (and the threshold for 
this instability is lowest) when the frequency of the pump beams is set near an 
atomic resonance, \emph{i.e.}, the instability occurs near either the D$_1$ or 
D$_2$ transition of $^{87}$Rb. For the remainder of the paper, we describe 
results for pump-beam frequencies near the D$_2$ transition ($^5\text{S}_{1/2} 
\rightarrow {^5\text{P}_{3/2}}$, 780 nm wavelength).

Figure~\ref{fig:output}(c) shows the power of the output light as a function
of pump frequency detuning, defined as $\Delta=\nu-\nu_{F=1,F'=1}$. We observe
several sub-Doppler features, where the maximum power emitted in the
orthogonal polarization occurs when the laser frequency $\nu$ is tuned
$\Delta=+25$~MHz. For this detuning, 3~$\mu$W of output light is generated,
indicating that $\sim 1$\% of the incident pump power is being converted to
the orthogonal polarization. Because the detuning is small relative to the
doppler width, most of the pump light is absorbed by the medium. Of the
80~$\mu$W of pump light transmitted in the forward direction, $\sim 4$\% is
being converted to the orthogonal polarization.


\begin{figure}
\includegraphics{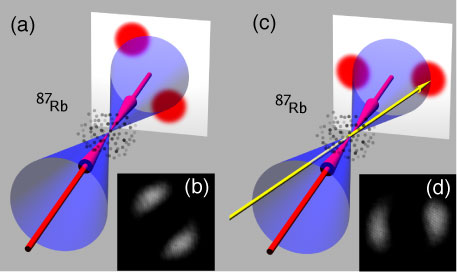}
\caption{\label{fig:cone_switch}(color online) Generated light is emitted along 
cones (blue), which are centered on the pump beams (red), and forms a 
transverse pattern (red spots) in the measurement plane. (a) Unperturbed system 
with two-spot pattern induced via weak pump-beam misalignment, (b) data showing 
two-spot pattern. (c) A weak switch beam (yellow) rotates the pattern, (d) data 
showing rotated pattern.}
\end{figure}

The light generated in our system shows structure in the transverse plane. We 
find that the output light is emitted in the forward and backward direction 
along cones centered on the pump beams (see Fig.~\ref{fig:cone_switch}(a)). The 
angle between the pump-beam axis and the cone is on the order of $\sim$5 mrad 
and can be understood as arising from competition between two different 
nonlinear processes: backward four-wave mixing in the phase-conjugation 
geometry and forward four-wave mixing \cite{Grynberg_1989aa}. 

We observe the cross-polarized off-axis output light on a measurement plane 
perpendicular to the propagation direction and in the far field.  
Figure~\ref{fig:output} shows a sample of the transverse patterns as a function 
of pump frequency detuning for pump powers above the instability-threshold. For 
pump beams that are red-detuned ($\Delta=-50$~MHz) the instability occurs 
on-axis, leading to a pattern containing one spot centered on the pump beam 
axis (Fig.~\ref{fig:output}(a)). For blue-detuned pump beams ($\Delta=25$~MHz) 
the instability occurs off-axis and the conical-emission is apparent from the 
hexagon pattern of the transverse field as shown in Fig.~\ref{fig:output}(b). 
Conical emission occurs only for blue-detuned beams where the sign of the 
nonlinear refractive index is positive \cite{Grynberg_1988ab, Firth_1988aa, 
Grynberg_1989aa}. The rotational symmetry, expected for conical emission, is 
spontaneously broken resulting in the formation of a hexagon pattern in the far 
field \cite{Geddes_1994aa}. The origin of the hexagonal pattern can be 
described by four-wave mixing where each spot on the hexagon is the result of a 
parametric process that annihilates one off-axis photon and one pump photon and 
creates two off-axis photons, both with wavevectors corresponding to the two 
next-nearest spots in the hexagon \cite{Grynberg_1988ab}. It should also be 
noted that the hexagonal symmetry is slightly broken by variations in the 
surface of the glass vapor cell and other experimental imperfections, even for 
well-aligned pump beams. This is manifested in the preferred azimuthal angle of 
emission: the upper-left and lower-right lobes of Fig.~\ref{fig:output}(b) are 
brighter than the other lobes.

This weak symmetry breaking can be made stronger by introducing a slight 
amount, less than 1~mrad, of mis-alignment between the pump beams. Doing so 
results in a two-spot pattern for pump beams just above threshold and detuned 
$\Delta=+25$~MHz (see Fig.~\ref{fig:cone_switch}(b)). The azimuthal orientation 
angle of these spots is stable for several minutes and is reproduced upon 
turning the pump beams off and on. The orientation can be influenced by the 
misalignment of the pump beams, application of a weak magnetic field or by 
injecting a weak probe beam, as discussed in Sec.~\ref{sec:switching}.



\begin{figure}
\includegraphics{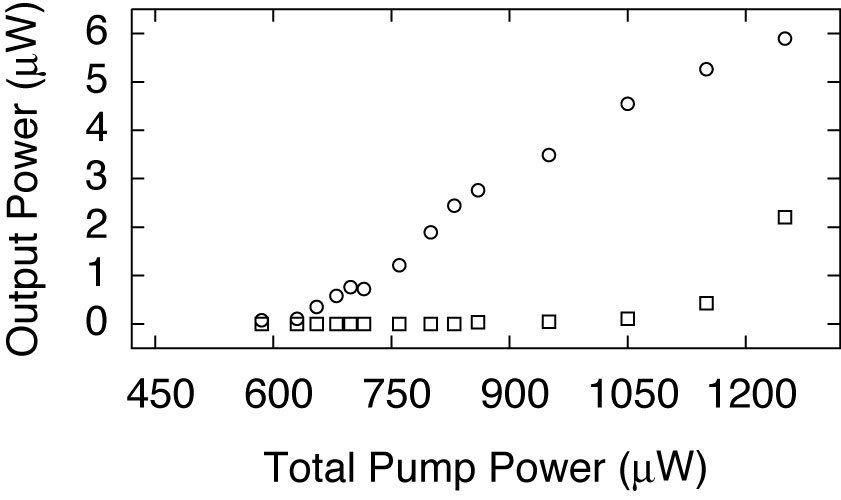}
\caption{\label{fig:threshold} Instability thresholds. Generated optical power 
as a function of total pump beam power. The off-axis instability occurs with a 
threshold of 600~$\mu$W (circles) whereas the on-axis instability occurs with a 
threshold of 1.1~mW (squares). Data shown corresponds to a fixed pump beam 
power ratio of 2.7 to 1 (forward to backward).}
\end{figure}

In addition to exhibiting pattern formation, the instability observed in our 
system has a very low threshold. The instability-threshold determines the 
lowest pump-power for which output light is generated. A common way to quantify 
the instability threshold for a setup with counterpropagating beams is to fix 
the power of one of the beams and measure the output power as a function of the 
power in the second pump beam \cite{Gauthier_1988aa, Zibrov_1999aa}. For a 
pump-beam detuning of $\Delta=+25$~MHz and with a fixed forward pump power of 
630~$\mu$W, we find that the backward pump power threshold is $\sim$125~$\mu$W, 
corresponding to a total pump power of 755~$\mu$W.

Another way to measure the instability threshold is to determine the minimum 
total pump power necessary to generate output light. Figure~\ref{fig:threshold} 
shows the result of such a measurement for our experiment. We find that there 
is an optimum ratio of forward power to backward power of $\sim$2.7-to-1. At 
this ratio, we determine the threshold for off-axis emission to be 600~$\mu$W 
which is slightly lower than the threshold measured with fixed forward beam 
power. Also shown in Fig.~\ref{fig:threshold}, for comparison, are data for the 
on-axis instability, detuning $\Delta=-50$~MHz, with a higher threshold of 
1.1~mW.

Both threshold measures demonstrate that the nonlinear process that generates 
new light is induced by a pair of very weak fields indicating very strong 
nonlinear matter-light interaction comparable with the best reported results to 
date for warm-vapor counterpropagating beam systems \cite{Zibrov_1999aa}. 

For most of the early observations of nearly-degenerate instabilities, strong 
pump fields were used (typically hundreds of mW) 
\cite{Grynberg_1988aa,Petrossian_1992aa,Maitre_1995aa}. A considerably higher 
threshold was reported in a previous work by one of us for polarization 
instabilities in a sodium vapor \cite{Gauthier_1988aa}, where a threshold of 
tens of mW was found when the pump fields were tuned near an atomic resonance. 
More recently, Zibrov {\em et al.}\ \cite{Zibrov_1999aa} observed parametric 
self-oscillation with pump powers in the $\mu$W regime using a more involved 
experimental setup (``double-$\Lambda$'' configuration) designed specifically 
to lower the instability threshold. In their experiment, atomic coherence 
effects increase the nonlinear coupling efficiency. They report oscillation 
with as little as 300~$\mu$W in the forward beam. With 5~mW of forward-beam 
power, their instability threshold corresponds to 20~$\mu$W in the backward 
beam. In contrast, the results reported here demonstrate that spontaneous 
parametric oscillations are induced by $\mu$W-power counterpropagating 
pump-beams without the need for special coherent preparation of the medium. 
Furthermore, Zibrov \emph{et al.}\ observe only on-axis emission, whereas, with 
our pump beam configuration, we find that off-axis emission requires roughly 
half as much pump power as on-axis emission. In situations where low power and 
high sensitivity are important, such as in all-optical switching, the lower 
instability threshold may make off-axis instabilities preferable.



\section{Switching Transverse Patterns} 
\label{sec:switching}

We find the azimuthal angle of the instability-generated beams is extremely 
sensitive to perturbations because the symmetry breaking of our setup is small. 
The preferred azimuthal orientation of the pattern can be overcome by injecting 
a weak switching beam along the cone of emission at a different azimuth as 
shown in Fig~\ref{fig:cone_switch}(c). Typically, this causes the pattern to 
rotate such that one spot is aligned to the switching beam with essentially no 
change in the total power of the pattern.

\begin{figure}
\includegraphics{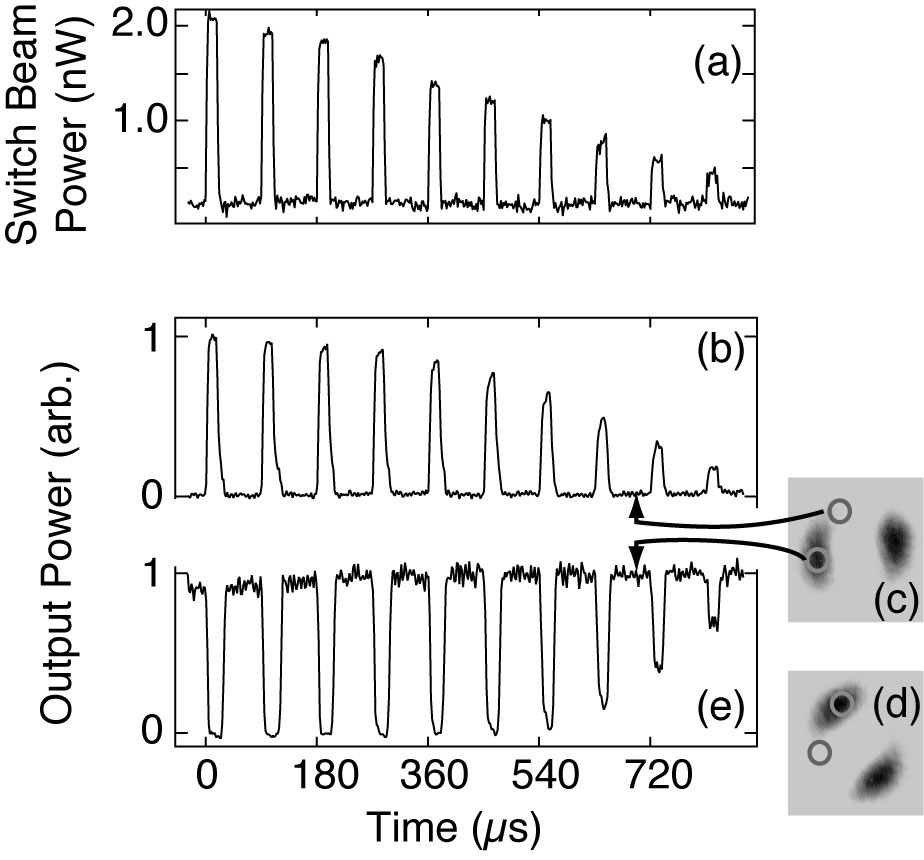}
\caption{\label{fig:spots_time} Time-response of the pattern. While turning the 
switch beam on and off with successively decreasing power levels (a), two 
detectors simultaneously measure the spatial regions indicated on the center 
frames (c,d) by circles which encompass bright and dark areas of the output 
pattern (the images have been inverted for visibility). These images are 
collected by CCD camera and the normalized detector output signals are shown in 
(b,e) corresponding to the spatial location of the detector within the pattern. 
(b) shows the ``on-state'' detector and (e) shows the ``off-state'' detector.}
\end{figure}

To quantify the dynamic behavior of the switch, we inject of series of pulses  
by turning the switch beam on and off with the Mach-Zehnder amplitude 
modulator. Spatially filtering the output pattern enables direct measurement of 
the switch behavior. High-contrast switching is confirmed by simultaneously 
measuring two output spots, one corresponding to a bright spot that exists in 
the absence of the switch beam (off state, lower circle in 
Fig~\ref{fig:spots_time}(c,d)) and the other corresponding to a spot that 
exists when the switch beam is present (on state, upper circle in 
Fig~\ref{fig:spots_time}(c,d)). The alternating signals shown in 
Fig~\ref{fig:spots_time}(b,e) demonstrate switching of the power from one 
switch state to another with high contrast. 

We inject switch beam pulses that steadily decrease in power, which allows us 
to sample the response of the system to various input levels. One notable 
feature of the system response is the transition from complete switching to 
partial switching. The first six pulses in Fig.~\ref{fig:spots_time} show that 
the on-state detector (Fig.~\ref{fig:spots_time}(b)) is fully illuminated and 
the off-state detector (Fig.~\ref{fig:spots_time}(e)) is dark. This indicates 
that the switch beam has caused complete rotation of the pattern and 
transferred all of the power from the off-state spots to the on-state spots. 
For the last four pulses in the series, the system exhibits partial switching, 
where the on-state detector is only partially illuminated and the off-state 
detector is partially darkened. This partial response indicates that the 
off-state spots are suppressed but not extinguished when the switch beam is 
applied with less than 1~nW. Similarly the on-state spots are generated but not 
at full power. In this intermediate regime, from 1~nW to 200~pW, the response 
is linearly proportional to the input power. Input powers below 200~pW do not 
rotate the spots so the output remains in the off state.

Also visible in Fig.~\ref{fig:spots_time}(e) is a weak secondary modulational 
instability that causes oscillations in the total output power. These 
oscillations are most visible in the off-state detector 
(Fig.~\ref{fig:spots_time}(e)) when the switch beam is not present. 
Experimental noise, primarily due to the detection electronics, is barely 
visible in this trace, hence the majority of the signal variation is due to 
this modulational instability. Fortunately, the effects of this secondary 
instability are minor, the primary effect being an increase in the switch 
response time uncertainty, as discussed below.


\section{Discussion} 
\label{sec:discussion}

\begin{figure}
\includegraphics{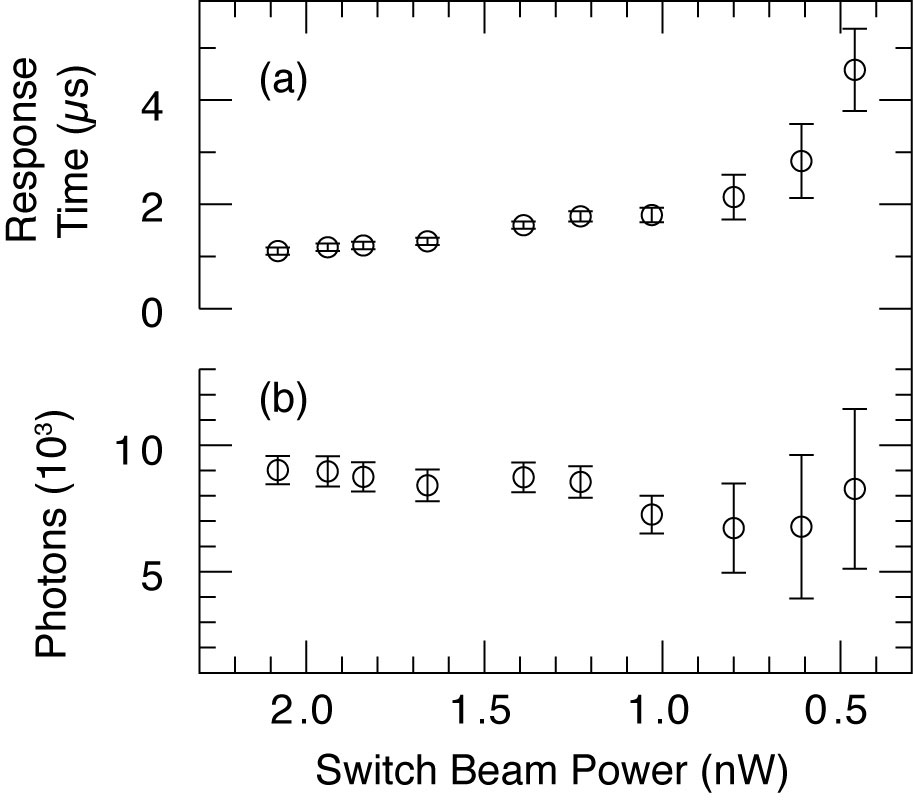}
\caption{\label{fig:photons} Response time and number of input photons as a 
function of switch beam power. (a) The response time increases for decreasing 
switch beam power. (b) the increased response time for weak input powers 
indicates that a constant number of photons are required to actuate the switch 
regardless of the amount of input power.}
\end{figure}

To quantify the sensitivity of the system, we measure the response time and 
from this calculate the number of photons required to actuate the switch. We 
define the response time of the device as the time between the initial rising 
edge of the switch beam pulse and the point where the on-spot signal crosses a 
threshold level set to correspond to a signal-to-noise ratio of 1. As shown in 
Fig.~\ref{fig:photons}(a), the measured response time increases as the input 
switch beam power decreases.

The number of photons required to actuate the switch is given by $N_p=\tau 
P_s/E_p$ where $\tau$ is the response time, $P_s$ is the switch beam power and 
$E_p=2.54\times10^{-19}$~J is the photon energy. For the ten switch-beam powers 
corresponding to Fig.~\ref{fig:spots_time}(a), the number of switching photons 
is plotted in Fig.~\ref{fig:photons}(b). The response time is longer for weak 
switch-beam powers so the photon number remains roughly constant, in the range 
between 6,000 and 9,000, regardless of the input power level. The noticeable 
increase in uncertainty for the latter points of Fig.~\ref{fig:photons}(a) is 
primarily due to the weak secondary modulational instability in the system. For 
weak input powers, the switch responds more quickly when the input pulse 
arrives in phase with the oscillations due to this instability. The error bars 
represent the range of response times observed. One goal of our future work is 
to minimize the effect of this secondary instability.

To compare the sensitivity of our switch to those previously discussed, we 
evaluate the energy density in photons per $\sigma$. For the switching beam 
spot size used, (1/e field radius) $w_0=235$~$\mu$m, 8,000 switching photons 
correspond to a switching beam energy density of $10^{-2}$ photons/$\sigma$. 
Therefore, both EIT-based switches \cite{Zhang_2007aa} and our pattern-based 
switch have surpassed the minimum value (1 photon/$\sigma$) originally expected 
for optical logic operations \cite{Keyes_1970aa}. Both approaches operate at 
very low light levels, although our system is markedly simpler than cold-atom 
EIT systems or cavity QED systems, requiring only one optical frequency and 
occurring in warm atomic vapor.

In addition to exhibiting high sensitivity, our switch is cascadable, with an 
output power capable of driving many subsequent devices. The 3~$\mu$W output 
power is sufficient to actuate over 1,000 similar switches, with each requiring 
$\sim$1~nW of input power.

The performance and scalability of a logic element can be further characterized 
by the device power requirements. In an early paper addressing the physical 
limits of logic devices, Keyes determined that optical logic based on the 
saturation of a two-level atomic transition requires input power levels that 
scale as $P\propto1/\tau^2$ where $P$ is the power required to complete one 
logic operation and $\tau$ is the response time of the device 
\cite{Keyes_1970aa}. This result leads to the conclusion that optical logic 
elements face thermal dissipation limits at lower speeds than electronic logic 
elements do, thus reducing their practical range of application.

A constant number of switching photons, however, implies that the power 
requirements of our device follows $P\propto 1/\tau$ rather than 
$P\propto1/\tau^2$. For response times below $10^{-8}$~s this is lower than the 
best-case estimate for an optical logic element based on saturating a two-level 
transition \cite{Keyes_1970aa}. This new scaling law for required power with 
response time indicates that the thermal dissipation limits previously assumed 
for optical logic devices are not as severe for our pattern-based switch and 
may extend the usability of optical logic to faster, more compact devices than 
was previously thought.

Finally, the two response regimes exhibited by our switch indicate that the 
output satisfies the conditions for signal level restoration. For a device to 
exhibit signal level restoration, variations in the input level cannot cause 
variations in the output level. In every device, however, there is a narrow 
range of input levels, known as the intermediate region, that lead to 
intermediate output levels. For input levels above or below the intermediate 
range the output is \emph{saturated} as a logic high or low respectively. In 
the case of our device, this intermediate region is between 200~pW and 1~nW. 
For input levels below 200 pW, the output is low, and for input levels above 1 
nW, the output is high with a level set by the pump beam power.

Signal level restoration is a key property of the electronic transistor and 
this demonstration of an optical logic element that exhibits level restoration 
is a key step towards practical optical switches. An all-optical transistor 
would have applications in many data processing and communication networks in 
the future.

\section{Conclusion} 
\label{sec:conclusion}

Instabilities that give rise to the generation of transverse optical patterns 
are induced in a simple atomic vapor system with less than 1 mW of pump power. 
These results suggest that coherently prepared media are not necessary for the 
observation of nonlinear optical interactions at low light levels. The 
sensitivity of these instabilities to perturbations enables incredibly weak 
optical fields to control the direction of much stronger beams generated by the 
instability. These patterns demonstrate low-light-level nonlinear optics using 
a simple setup and enable a new kind of all-optical switch that satisfies the 
requirements of scalability.

Other recent all-optical switches are based on quantum interference or cavity 
QED and require complex experimental environments to operate with a sensitivity 
of less than one photon per atomic cross section. Furthermore, previous devices 
fail to meet one or more of the criteria for scalability. We have extended our 
previous demonstration of a cascadable, pattern-based all-optical switch that 
operates at ultra-low light levels. The results presented here show that such a 
switch exhibits the transistor-like behavior of signal level restoration, and 
is operated in a very simple experimental environment with $<$1~mW of total 
pump power.


\begin{acknowledgments} We gratefully acknowledge the financial support of the 
U.S. Army Research Office Grant No. W911NF-05-1-0228, and the DARPA DSO 
Slow-Light Program. \end{acknowledgments}



\begin{thebibliography}{36}
  \expandafter\ifx\csname natexlab\endcsname\relax\def\natexlab#1{#1}\fi
  \expandafter\ifx\csname bibnamefont\endcsname\relax
    \def\bibnamefont#1{#1}\fi
  \expandafter\ifx\csname bibfnamefont\endcsname\relax
    \def\bibfnamefont#1{#1}\fi
  \expandafter\ifx\csname citenamefont\endcsname\relax
    \def\citenamefont#1{#1}\fi
  \expandafter\ifx\csname url\endcsname\relax
    \def\url#1{\texttt{#1}}\fi
  \expandafter\ifx\csname urlprefix\endcsname\relax\def\urlprefix{URL }\fi
  \providecommand{\bibinfo}[2]{#2}
  \providecommand{\eprint}[2][]{\url{#2}}

  \bibitem[{\citenamefont{Gibbs}(1985)}]{Gibbs_1985aa}
  \bibinfo{author}{\bibfnamefont{H.~M.} \bibnamefont{Gibbs}},
    \emph{\bibinfo{title}{Optical Bistability: Controlling Light with Light}}
    (\bibinfo{publisher}{Academic Press}, \bibinfo{address}{Orlando},
    \bibinfo{year}{1985}).

  \bibitem[{\citenamefont{Keyes}(2006)}]{Keyes_2006aa}
  \bibinfo{author}{\bibfnamefont{R.~W.} \bibnamefont{Keyes}},
    \bibinfo{journal}{J. Phys.: Condens. Matter} \textbf{\bibinfo{volume}{18}},
    \bibinfo{pages}{S703} (\bibinfo{year}{2006}).

  \bibitem[{\citenamefont{Bouwmeester et~al.}(2000)\citenamefont{Bouwmeester,
    Ekert, and Zeilinger}}]{Bouwmeester_2000aa}
  \bibinfo{author}{\bibfnamefont{D.}~\bibnamefont{Bouwmeester}},
    \bibinfo{author}{\bibfnamefont{A.~K.} \bibnamefont{Ekert}}, 
\bibnamefont{and}
    \bibinfo{author}{\bibfnamefont{A.}~\bibnamefont{Zeilinger}},
    \emph{\bibinfo{title}{The Physics of Quantum Information: Quantum
    Cryptography, Quantum Teleportation, Quantum Computation}}
    (\bibinfo{publisher}{Springer, Berlin}, \bibinfo{year}{2000}).

  \bibitem[{\citenamefont{Dawes et~al.}(2005)\citenamefont{Dawes, Illing, Clark,
    and Gauthier}}]{Dawes_2005aa}
  \bibinfo{author}{\bibfnamefont{A.~M.~C.} \bibnamefont{Dawes}},
    \bibinfo{author}{\bibfnamefont{L.}~\bibnamefont{Illing}},
    \bibinfo{author}{\bibfnamefont{S.~M.} \bibnamefont{Clark}}, 
\bibnamefont{and}
    \bibinfo{author}{\bibfnamefont{D.~J.} \bibnamefont{Gauthier}},
    \bibinfo{journal}{Science} \textbf{\bibinfo{volume}{308}},
    \bibinfo{pages}{672} (\bibinfo{year}{2005}).

  \bibitem[{\citenamefont{Harris and Yamamoto}(1998)}]{Harris_1998aa}
  \bibinfo{author}{\bibfnamefont{S.~E.} \bibnamefont{Harris}} \bibnamefont{and}
    \bibinfo{author}{\bibfnamefont{Y.}~\bibnamefont{Yamamoto}},
    \bibinfo{journal}{Phys. Rev. Lett.} \textbf{\bibinfo{volume}{81}},
    \bibinfo{pages}{3611} (\bibinfo{year}{1998}).

  \bibitem[{\citenamefont{Keyes}(1970)}]{Keyes_1970aa}
  \bibinfo{author}{\bibfnamefont{R.~W.} \bibnamefont{Keyes}},
    \bibinfo{journal}{Science} \textbf{\bibinfo{volume}{168}},
    \bibinfo{pages}{796} (\bibinfo{year}{1970}).

  \bibitem[{\citenamefont{Hood et~al.}(1998)\citenamefont{Hood, Chapman, Lynn,
    and Kimble}}]{Hood_1998aa}
  \bibinfo{author}{\bibfnamefont{C.~J.} \bibnamefont{Hood}},
    \bibinfo{author}{\bibfnamefont{M.~S.} \bibnamefont{Chapman}},
    \bibinfo{author}{\bibfnamefont{T.~W.} \bibnamefont{Lynn}}, \bibnamefont{and}
    \bibinfo{author}{\bibfnamefont{H.~J.} \bibnamefont{Kimble}},
    \bibinfo{journal}{Phys. Rev. Lett.} \textbf{\bibinfo{volume}{80}},
    \bibinfo{pages}{4157} (\bibinfo{year}{1998}).

  \bibitem[{\citenamefont{Birnbaum et~al.}(2005)\citenamefont{Birnbaum, Boca,
    Miller, Boozer, Northup, and Kimble}}]{Birnbaum_2005aa}
  \bibinfo{author}{\bibfnamefont{K.}~\bibnamefont{Birnbaum}},
    \bibinfo{author}{\bibfnamefont{A.}~\bibnamefont{Boca}},
    \bibinfo{author}{\bibfnamefont{R.}~\bibnamefont{Miller}},
    \bibinfo{author}{\bibfnamefont{A.}~\bibnamefont{Boozer}},
    \bibinfo{author}{\bibfnamefont{T.}~\bibnamefont{Northup}}, \bibnamefont{and}
    \bibinfo{author}{\bibfnamefont{H.}~\bibnamefont{Kimble}},
    \bibinfo{journal}{Nature} \textbf{\bibinfo{volume}{436}}, 
\bibinfo{pages}{87}
    (\bibinfo{year}{2005}).

  \bibitem[{\citenamefont{Zhang et~al.}(2007)\citenamefont{Zhang, Hernandez, and
    Zhu}}]{Zhang_2007aa}
  \bibinfo{author}{\bibfnamefont{J.}~\bibnamefont{Zhang}},
    \bibinfo{author}{\bibfnamefont{G.}~\bibnamefont{Hernandez}},
    \bibnamefont{and} \bibinfo{author}{\bibfnamefont{Y.}~\bibnamefont{Zhu}},
    \bibinfo{journal}{Opt. Lett.} \textbf{\bibinfo{volume}{32}},
    \bibinfo{pages}{1317} (\bibinfo{year}{2007}).

  \bibitem[{\citenamefont{Hachair et~al.}(2005)\citenamefont{Hachair, Furfaro,
    Javaloyes, Giudici, Balle, Tredicce, Tissoni, Lugiato, Brambilla, and
    Maggipinto}}]{Hachair_2005aa}
  \bibinfo{author}{\bibfnamefont{X.}~\bibnamefont{Hachair}},
    \bibinfo{author}{\bibfnamefont{L.}~\bibnamefont{Furfaro}},
    \bibinfo{author}{\bibfnamefont{J.}~\bibnamefont{Javaloyes}},
    \bibinfo{author}{\bibfnamefont{M.}~\bibnamefont{Giudici}},
    \bibinfo{author}{\bibfnamefont{S.}~\bibnamefont{Balle}},
    \bibinfo{author}{\bibfnamefont{J.}~\bibnamefont{Tredicce}},
    \bibinfo{author}{\bibfnamefont{G.}~\bibnamefont{Tissoni}},
    \bibinfo{author}{\bibfnamefont{L.~A.} \bibnamefont{Lugiato}},
    \bibinfo{author}{\bibfnamefont{M.}~\bibnamefont{Brambilla}},
    \bibnamefont{and}
    \bibinfo{author}{\bibfnamefont{T.}~\bibnamefont{Maggipinto}},
    \bibinfo{journal}{Phys. Rev. A} \textbf{\bibinfo{volume}{72}},
    \bibinfo{eid}{013815} (\bibinfo{year}{2005}).

  \bibitem[{\citenamefont{Harris}(1997)}]{Harris_1997aa}
  \bibinfo{author}{\bibfnamefont{S.~E.} \bibnamefont{Harris}},
    \bibinfo{journal}{Phys. Today} \textbf{\bibinfo{volume}{50}},
    \bibinfo{pages}{36} (\bibinfo{year}{1997}).

  \bibitem[{\citenamefont{Schmidt and Imamoglu}(1996)}]{Schmidt_1996aa}
  \bibinfo{author}{\bibfnamefont{H.}~\bibnamefont{Schmidt}} \bibnamefont{and}
    \bibinfo{author}{\bibfnamefont{A.}~\bibnamefont{Imamoglu}},
    \bibinfo{journal}{Opt. Lett.} \textbf{\bibinfo{volume}{21}},
    \bibinfo{pages}{1936} (\bibinfo{year}{1996}).

  \bibitem[{\citenamefont{Zibrov et~al.}(1999)\citenamefont{Zibrov, Lukin, and
    Scully}}]{Zibrov_1999aa}
  \bibinfo{author}{\bibfnamefont{A.~S.} \bibnamefont{Zibrov}},
    \bibinfo{author}{\bibfnamefont{M.~D.} \bibnamefont{Lukin}}, 
\bibnamefont{and}
    \bibinfo{author}{\bibfnamefont{M.~O.} \bibnamefont{Scully}},
    \bibinfo{journal}{Phys. Rev. Lett.} \textbf{\bibinfo{volume}{83}},
    \bibinfo{pages}{4049} (\bibinfo{year}{1999}).

  \bibitem[{\citenamefont{Chen et~al.}(2005)\citenamefont{Chen, Tsai, Liu, and
    Yu}}]{Chen_2005aa}
  \bibinfo{author}{\bibfnamefont{Y.-F.} \bibnamefont{Chen}},
    \bibinfo{author}{\bibfnamefont{Z.-H.} \bibnamefont{Tsai}},
    \bibinfo{author}{\bibfnamefont{Y.-C.} \bibnamefont{Liu}}, \bibnamefont{and}
    \bibinfo{author}{\bibfnamefont{I.~A.} \bibnamefont{Yu}},
    \bibinfo{journal}{Opt. Lett.} \textbf{\bibinfo{volume}{30}},
    \bibinfo{pages}{3207} (\bibinfo{year}{2005}).

  \bibitem[{\citenamefont{Braje et~al.}(2003)\citenamefont{Braje, Bali{\'c}, 
Yin,
    and Harris}}]{Braje_2003aa}
  \bibinfo{author}{\bibfnamefont{D.~A.} \bibnamefont{Braje}},
    \bibinfo{author}{\bibfnamefont{V.}~\bibnamefont{Bali{\'c}}},
    \bibinfo{author}{\bibfnamefont{G.~Y.} \bibnamefont{Yin}}, \bibnamefont{and}
    \bibinfo{author}{\bibfnamefont{S.~E.} \bibnamefont{Harris}},
    \bibinfo{journal}{Phys. Rev. A} \textbf{\bibinfo{volume}{68}},
    \bibinfo{pages}{041801(R)} (\bibinfo{year}{2003}).

  \bibitem[{\citenamefont{Resch et~al.}(2002)\citenamefont{Resch, Lundeen, and
    Steinberg}}]{Resch_2002aa}
  \bibinfo{author}{\bibfnamefont{K.~J.} \bibnamefont{Resch}},
    \bibinfo{author}{\bibfnamefont{J.~S.} \bibnamefont{Lundeen}},
    \bibnamefont{and} \bibinfo{author}{\bibfnamefont{A.~M.}
    \bibnamefont{Steinberg}}, \bibinfo{journal}{Phys. Rev. Lett.}
    \textbf{\bibinfo{volume}{89}}, \bibinfo{pages}{037904}
    (\bibinfo{year}{2002}).

  \bibitem[{\citenamefont{Kang et~al.}(2004)\citenamefont{Kang, Hernandez, and
    Zhu}}]{Kang_2004aa}
  \bibinfo{author}{\bibfnamefont{H.}~\bibnamefont{Kang}},
    \bibinfo{author}{\bibfnamefont{G.}~\bibnamefont{Hernandez}},
    \bibnamefont{and} \bibinfo{author}{\bibfnamefont{Y.}~\bibnamefont{Zhu}},
    \bibinfo{journal}{Phys. Rev. Lett.} \textbf{\bibinfo{volume}{93}},
    \bibinfo{pages}{073601} (\bibinfo{year}{2004}).

  \bibitem[{\citenamefont{Wang et~al.}(2002)\citenamefont{Wang, Goorskey, and
    Xiao}}]{Wang_2002aa}
  \bibinfo{author}{\bibfnamefont{H.}~\bibnamefont{Wang}},
    \bibinfo{author}{\bibfnamefont{D.}~\bibnamefont{Goorskey}}, 
\bibnamefont{and}
    \bibinfo{author}{\bibfnamefont{M.}~\bibnamefont{Xiao}},
    \bibinfo{journal}{Phys. Rev. A} \textbf{\bibinfo{volume}{65}},
    \bibinfo{pages}{051802(R)} (\bibinfo{year}{2002}).

  \bibitem[{\citenamefont{Tanabe et~al.}(2005)\citenamefont{Tanabe, Notomi,
    Mitsugi, Shinya, and Kuramochi}}]{Tanabe_2005aa}
  \bibinfo{author}{\bibfnamefont{T.}~\bibnamefont{Tanabe}},
    \bibinfo{author}{\bibfnamefont{M.}~\bibnamefont{Notomi}},
    \bibinfo{author}{\bibfnamefont{S.}~\bibnamefont{Mitsugi}},
    \bibinfo{author}{\bibfnamefont{A.}~\bibnamefont{Shinya}}, \bibnamefont{and}
    \bibinfo{author}{\bibfnamefont{E.}~\bibnamefont{Kuramochi}},
    \bibinfo{journal}{Opt. Lett.} \textbf{\bibinfo{volume}{30}},
    \bibinfo{pages}{2575} (\bibinfo{year}{2005}).

  \bibitem[{\citenamefont{Solja{\v c}i{\'c} et~al.}(2005)\citenamefont{Solja{\v
    c}i{\'c}, Lidorikis, Joannopoulos, and Hau}}]{Soljacic_2005aa}
  \bibinfo{author}{\bibfnamefont{M.}~\bibnamefont{Solja{\v c}i{\'c}}},
    \bibinfo{author}{\bibfnamefont{E.}~\bibnamefont{Lidorikis}},
    \bibinfo{author}{\bibfnamefont{J.~D.} \bibnamefont{Joannopoulos}},
    \bibnamefont{and} \bibinfo{author}{\bibfnamefont{L.~V.} \bibnamefont{Hau}},
    \bibinfo{journal}{Appl. Phys. Lett.} \textbf{\bibinfo{volume}{86}},
    \bibinfo{eid}{171101} (\bibinfo{year}{2005}).

  \bibitem[{\citenamefont{Islam et~al.}(1988)\citenamefont{Islam, Dijaili, and
    Gordon}}]{Islam_1988aa}
  \bibinfo{author}{\bibfnamefont{M.}~\bibnamefont{Islam}},
    \bibinfo{author}{\bibfnamefont{S.}~\bibnamefont{Dijaili}}, \bibnamefont{and}
    \bibinfo{author}{\bibfnamefont{J.}~\bibnamefont{Gordon}},
    \bibinfo{journal}{Opt. Lett.} \textbf{\bibinfo{volume}{13}},
    \bibinfo{pages}{518} (\bibinfo{year}{1988}).

  \bibitem[{\citenamefont{Chang et~al.}(2007)\citenamefont{Chang, Sorensen,
    Demler, and Lukin}}]{Chang_2007ab}
  \bibinfo{author}{\bibfnamefont{D.~E.} \bibnamefont{Chang}},
    \bibinfo{author}{\bibfnamefont{A.~S.} \bibnamefont{Sorensen}},
    \bibinfo{author}{\bibfnamefont{E.~A.} \bibnamefont{Demler}},
    \bibnamefont{and} \bibinfo{author}{\bibfnamefont{M.~D.} 
\bibnamefont{Lukin}},
    \bibinfo{journal}{Nature Physics, advance online publication, 26 August 2007
    (doi:10.1038/nphys708)}  (\bibinfo{year}{2007}).

  \bibitem[{\citenamefont{Lugiato}(1994)}]{Lugiato_1994aa}
  \bibinfo{author}{\bibfnamefont{L.~A.} \bibnamefont{Lugiato}},
    \bibinfo{journal}{Chaos, Solitons, \& Fractals} 
\textbf{\bibinfo{volume}{4}},
    \bibinfo{pages}{1251} (\bibinfo{year}{1994}).

  \bibitem[{\citenamefont{Silberberg and Bar-Joseph}(1984)}]{Silberberg_1984aa}
  \bibinfo{author}{\bibfnamefont{Y.}~\bibnamefont{Silberberg}} \bibnamefont{and}
    \bibinfo{author}{\bibfnamefont{I.}~\bibnamefont{Bar-Joseph}},
    \bibinfo{journal}{J. Opt. Soc. Am. B} \textbf{\bibinfo{volume}{1}},
    \bibinfo{pages}{662} (\bibinfo{year}{1984}).

  \bibitem[{\citenamefont{Khitrova et~al.}(1988)\citenamefont{Khitrova, Valley,
    and Gibbs}}]{Khitrova_1988aa}
  \bibinfo{author}{\bibfnamefont{G.}~\bibnamefont{Khitrova}},
    \bibinfo{author}{\bibfnamefont{J.~F.} \bibnamefont{Valley}},
    \bibnamefont{and} \bibinfo{author}{\bibfnamefont{H.~M.} 
\bibnamefont{Gibbs}},
    \bibinfo{journal}{Phys. Rev. Lett.} \textbf{\bibinfo{volume}{60}},
    \bibinfo{pages}{1126} (\bibinfo{year}{1988}).

  \bibitem[{\citenamefont{Gauthier et~al.}(1990)\citenamefont{Gauthier, Malcuit,
    Gaeta, and Boyd}}]{Gauthier_1990aa}
  \bibinfo{author}{\bibfnamefont{D.~J.} \bibnamefont{Gauthier}},
    \bibinfo{author}{\bibfnamefont{M.~S.} \bibnamefont{Malcuit}},
    \bibinfo{author}{\bibfnamefont{A.~L.} \bibnamefont{Gaeta}}, 
\bibnamefont{and}
    \bibinfo{author}{\bibfnamefont{R.~W.} \bibnamefont{Boyd}},
    \bibinfo{journal}{Phys. Rev. Lett.} \textbf{\bibinfo{volume}{64}},
    \bibinfo{pages}{1721} (\bibinfo{year}{1990}).

  \bibitem[{\citenamefont{Gauthier et~al.}(1988)\citenamefont{Gauthier, Malcuit,
    and Boyd}}]{Gauthier_1988aa}
  \bibinfo{author}{\bibfnamefont{D.~J.} \bibnamefont{Gauthier}},
    \bibinfo{author}{\bibfnamefont{M.~S.} \bibnamefont{Malcuit}},
    \bibnamefont{and} \bibinfo{author}{\bibfnamefont{R.~W.} \bibnamefont{Boyd}},
    \bibinfo{journal}{Phys. Rev. Lett.} \textbf{\bibinfo{volume}{61}},
    \bibinfo{pages}{1827} (\bibinfo{year}{1988}).

  \bibitem[{\citenamefont{Gaeta et~al.}(1987)\citenamefont{Gaeta, Boyd,
    Ackerhalt, and Milonni}}]{Gaeta_1987aa}
  \bibinfo{author}{\bibfnamefont{A.~L.} \bibnamefont{Gaeta}},
    \bibinfo{author}{\bibfnamefont{R.~W.} \bibnamefont{Boyd}},
    \bibinfo{author}{\bibfnamefont{J.~R.} \bibnamefont{Ackerhalt}},
    \bibnamefont{and} \bibinfo{author}{\bibfnamefont{P.~W.}
    \bibnamefont{Milonni}}, \bibinfo{journal}{Phys. Rev. Lett.}
    \textbf{\bibinfo{volume}{58}}, \bibinfo{pages}{2432 } 
(\bibinfo{year}{1987}).

  \bibitem[{\citenamefont{Petrossian et~al.}(1992)\citenamefont{Petrossian,
    Pinard, Ma{\^\i}tre, Courtois, and Grynberg}}]{Petrossian_1992aa}
  \bibinfo{author}{\bibfnamefont{A.}~\bibnamefont{Petrossian}},
    \bibinfo{author}{\bibfnamefont{M.}~\bibnamefont{Pinard}},
    \bibinfo{author}{\bibfnamefont{A.}~\bibnamefont{Ma{\^\i}tre}},
    \bibinfo{author}{\bibfnamefont{J.~Y.} \bibnamefont{Courtois}},
    \bibnamefont{and} 
\bibinfo{author}{\bibfnamefont{G.}~\bibnamefont{Grynberg}},
    \bibinfo{journal}{Europhys. Lett.} \textbf{\bibinfo{volume}{18}},
    \bibinfo{pages}{689 } (\bibinfo{year}{1992}).

  \bibitem[{\citenamefont{Grynberg et~al.}(1988)\citenamefont{Grynberg, 
Le~Bihan,
    Verklerk, Simoneau, Leite, Bloch, Le~Boiteux, and Ducloy}}]{Grynberg_1988aa}
  \bibinfo{author}{\bibfnamefont{G.}~\bibnamefont{Grynberg}},
    \bibinfo{author}{\bibfnamefont{E.}~\bibnamefont{Le~Bihan}},
    \bibinfo{author}{\bibfnamefont{P.}~\bibnamefont{Verklerk}},
    \bibinfo{author}{\bibfnamefont{P.}~\bibnamefont{Simoneau}},
    \bibinfo{author}{\bibfnamefont{J.~R.~R.} \bibnamefont{Leite}},
    \bibinfo{author}{\bibfnamefont{D.}~\bibnamefont{Bloch}},
    \bibinfo{author}{\bibfnamefont{S.}~\bibnamefont{Le~Boiteux}},
    \bibnamefont{and} \bibinfo{author}{\bibfnamefont{M.}~\bibnamefont{Ducloy}},
    \bibinfo{journal}{Opt. Commun.} \textbf{\bibinfo{volume}{67}},
    \bibinfo{pages}{363} (\bibinfo{year}{1988}).

  \bibitem[{\citenamefont{Gaeta and Boyd}(1993)}]{Gaeta_1993aa}
  \bibinfo{author}{\bibfnamefont{A.~L.} \bibnamefont{Gaeta}} \bibnamefont{and}
    \bibinfo{author}{\bibfnamefont{R.~W.} \bibnamefont{Boyd}},
    \bibinfo{journal}{Phys. Rev. A} \textbf{\bibinfo{volume}{48}},
    \bibinfo{pages}{1610} (\bibinfo{year}{1993}).

  \bibitem[{\citenamefont{Grynberg and Paye}(1989)}]{Grynberg_1989aa}
  \bibinfo{author}{\bibfnamefont{G.}~\bibnamefont{Grynberg}} \bibnamefont{and}
    \bibinfo{author}{\bibfnamefont{J.}~\bibnamefont{Paye}},
    \bibinfo{journal}{Europhys. Lett.} \textbf{\bibinfo{volume}{8}},
    \bibinfo{pages}{29 } (\bibinfo{year}{1989}).

  \bibitem[{\citenamefont{Grynberg}(1988)}]{Grynberg_1988ab}
  \bibinfo{author}{\bibfnamefont{G.}~\bibnamefont{Grynberg}},
    \bibinfo{journal}{Opt. Commun.} \textbf{\bibinfo{volume}{66}},
    \bibinfo{pages}{321} (\bibinfo{year}{1988}).

  \bibitem[{\citenamefont{Firth and Par{\'e}}(1988)}]{Firth_1988aa}
  \bibinfo{author}{\bibfnamefont{W.~J.} \bibnamefont{Firth}} \bibnamefont{and}
    \bibinfo{author}{\bibfnamefont{C.}~\bibnamefont{Par{\'e}}},
    \bibinfo{journal}{Opt. Lett.} \textbf{\bibinfo{volume}{13}},
    \bibinfo{pages}{1096} (\bibinfo{year}{1988}).

  \bibitem[{\citenamefont{Geddes et~al.}(1994)\citenamefont{Geddes, Indik,
    Moloney, and Firth}}]{Geddes_1994aa}
  \bibinfo{author}{\bibfnamefont{J.~B.} \bibnamefont{Geddes}},
    \bibinfo{author}{\bibfnamefont{R.~A.} \bibnamefont{Indik}},
    \bibinfo{author}{\bibfnamefont{J.~V.} \bibnamefont{Moloney}},
    \bibnamefont{and} \bibinfo{author}{\bibfnamefont{W.~J.} 
\bibnamefont{Firth}},
    \bibinfo{journal}{Phys. Rev. A} \textbf{\bibinfo{volume}{50}},
    \bibinfo{pages}{3471} (\bibinfo{year}{1994}).

  \bibitem[{\citenamefont{Ma{\^\i}tre et~al.}(1995)\citenamefont{Ma{\^\i}tre,
    Petrossian, Blouin, Pinard, and Grynberg}}]{Maitre_1995aa}
  \bibinfo{author}{\bibfnamefont{A.}~\bibnamefont{Ma{\^\i}tre}},
    \bibinfo{author}{\bibfnamefont{A.}~\bibnamefont{Petrossian}},
    \bibinfo{author}{\bibfnamefont{A.}~\bibnamefont{Blouin}},
    \bibinfo{author}{\bibfnamefont{M.}~\bibnamefont{Pinard}}, \bibnamefont{and}
    \bibinfo{author}{\bibfnamefont{G.}~\bibnamefont{Grynberg}},
    \bibinfo{journal}{Opt. Commun.} \textbf{\bibinfo{volume}{116}},
    \bibinfo{pages}{153} (\bibinfo{year}{1995}).

  \end{thebibliography}
\end{document}